\documentclass[sigplan]{acmart}
\pdfoutput=1 

\usepackage{algpseudocode,algorithm,algorithmicx}
\usepackage{amsthm}
\usepackage{booktabs} 
\usepackage{graphicx}
\usepackage[normalem]{ulem}
\usepackage{listings}
\usepackage{xspace}
\usepackage[skip=2pt]{caption} 

\makeatletter
\def\BState{\State\hskip-\ALG@thistlm}
\makeatother

\newcommand{\jnote}[1]{{\color{red} Jim: #1}}
\newcommand{\nnote}[1]{{\color{violet} Nachshon: #1}}

\renewcommand{\jnote}[1]{}
\renewcommand{\nnote}[1]{}

\newcommand{\ignore}[1]{}
\newcommand{\desc}[1]{}

\newcommand{\nanosecond}{\(ns\)\xspace}
\newcommand{\microsecond}{\(\upmu{}\)s\xspace}
\newcommand{\millisecond}{\(ms\)\xspace}

\copyrightyear{2019}
\acmYear{2019}
\setcopyright{rightsretained}
\acmConference[ASPLOS '19]{2019 Architectural Support for Programming
	Languages and Operating Systems}{April 13--17, 2019}{Providence, RI, USA}
\acmBooktitle{2019 Architectural Support for Programming Languages and
	Operating Systems (ASPLOS '19), April 13--17, 2019, Providence, RI, USA}
\acmDOI{10.1145/3297858.3304046}
\acmISBN{978-1-4503-6240-5/19/04}

\begin{document}
\title{Fine-Grain Checkpointing with In-Cache-Line Logging}

\author{Nachshon Cohen}
\orcid{0000-0001-8302-2739}             
\affiliation{
	\institution{Amazon\thanks{Work done while the first author was a postdoc at EPFL}}            
	\city{Haifa}
	\country{Israel}
}
\email{nachshonc@gmail.com}          

\author{David T. Aksun}
\affiliation{
	\institution{EPFL}            
	\city{Lausanne}
	\country{Switzerland}
}
\email{david.aksun@epfl.ch}          

\author{Hillel Avni}
\affiliation{
	\institution{Huawei}            
	\city{Tel Aviv}
	\country{Israel}
}
\email{hillel.avni@huawei.com}

\author{James R. Larus}
\affiliation{
\institution{EPFL}            
	\city{Lausanne}
	\country{Switzerland}
}
\email{james.larus@epfl.ch}          

\date{}

\begin{abstract}
	Non-Volatile Memory offers the possibility of implementing high-performance, durable data structures.
However, achieving performance comparable to well-designed data structures in non-persistent (transient) memory is difficult, primarily because of the cost of ensuring the order in which memory writes reach NVM.\@
Often, this requires flushing data to NVM and waiting a full memory round-trip time.

In this paper, we introduce two new techniques: \emph{Fine-Grained Checkpointing}, which ensures a consistent, quickly recoverable data structure in NVM after a system failure, and \emph{In-Cache-Line Logging}, an undo-logging technique that enables recovery of earlier state without requiring cache-line flushes in the normal case.
We implemented these techniques in the Masstree data structure, making it persistent and demonstrating the ease of applying them to a highly optimized system and their low (5.9-15.4\%) runtime overhead cost. 
\end{abstract}


\maketitle

\section{Introduction}
Non-Volatile Memory (NVM) is fast, byte-addressable memory that retains its contents after a power failure or a system crash. 
New technologies, such as 3D-XPoint~\cite{Intel2015}, PCM~\cite{Qureshi2009,Lee2009}, STT-RAM~\cite{Hosomi}, and ReRAM~\cite{Akinaga2010,Wong2012}, promise NVM at low cost, thus blurring the line between durable storage and main memory.
One important use of NVM is enabling the rapid restart of a failed system.
Restarting an existing machine typically incurs a significant delay due to the need to read data from durable media such as a disk or SSD, parse it, and rebuild internal data structures.
NVM can avoid most of these restart costs. 
Since NVM is byte-addressable, it is possible to store efficient, pointer-based structures, such as B+ trees or hash maps, directly in NVM.\@ 
After a failure, these structures remain in NVM, enabling the system to resume immediately after rebooting and recovering data~\cite{Cohen2018}. 

The main challenge in designing durable data structures for NVM is that processor caches are (and are likely to remain) transient. 
During a power failure, all memory writes that were not propagated from cache to NVM will be lost.
The processor memory system also complicates the task of writing cache lines to NVM in a consistent manner.
Cache lines are not written back to memory (NVM) in the order in which an application modifies them, but rather according to a memory system's low-level and frequently undocumented cache replacement policy. 
This creates a well-studied challenge: how to ensure that the durable copy of a data structure is well-formed (consistent) after a crash, even though NVM contains a mixture of stale and new cache lines?

Most NVM systems require programmer-specified transactions to ensure that a group of memory writes either all reach NVM or none of them do.
Typically, a change to a structure is first logged (using either a redo or undo log) in NVM and then applied to the structure.
The log provides sufficient information for the recovery process to restore a structure to a consistent state, regardless of whether all structure modifications reached NVM.\@

However, the system must ensure that the log is completely flushed to NVM before the structure itself is modified.
This requires the use of cache flush instructions that transfer dirty cache lines from the (transient) cache to the (durable) NVM.\@
These write backs are only guaranteed to have completed after a fence instruction executes.
These instructions are expensive since they require a full round trip to NVM and reduce program performance by a significant amount.
This overhead, moreover, can be incurred on an application's critical path, as a structure is being updated.

An alternative to transactions is checkpointing, another widely used technique for ensuring recoverability. 
At periodic intervals, an application's entire state is saved on durable media.
After a failure, the last recorded checkpoint is restored to memory and the computation resumes from this point, requiring the re-execution of the work done between the checkpoint and failure.
Since copying the entire state of an application to slow durable media is expensive, most systems take checkpoints at infrequent intervals (minutes to hours) to reduce this cost.
The interval between checkpoints is a tradeoff between the overhead of recording checkpoints and the cost of re-executing the lost computation.

In this work, we introduce \emph{Fine-Grained Checkpointing}, which uses frequent checkpoints to NVM to ensure persistence at low cost.
Instead of ensuring that every memory write is logged or propagates to NVM, we partition an application's execution into epochs and ensure that after a crash, data structures can be restored to their state at the end of the most recently completed epoch (Nawab calls a similar approach ``periodic persistence''~\cite{Nawab2017}).\footnote{Unlike most checkpointing, which resumes execution at the point at which a checkpoint occurred, our goal is to restore the persistent data structures to their state at the checkpoint, so that a program can \emph{restart} its execution.}
Our system flushes the processors' caches to NVM at the start of an epoch, thus ensuring that at this point NVM contains \emph{all} modified data.

This approach has many advantages.
The number of cache lines that must be flushed is bounded by the cache size, and modified cache lines may have been written back during the epoch, so the cost of recording a checkpoint is low.
The modified lines are flushed in a batch by hardware, further reducing the cost. 
Our approach's low cost allows short checkpoint intervals, e.g., tens of milliseconds (we use 64\millisecond), thereby reducing both the potential data loss and the recovery time.
In addition, a software developer need not annotate an application to delimit fine-grained transactions, rather he or she only needs to ensure that the application's state is recoverable at the end of an epoch.

Our approach differs from a traditional checkpoint in that there is no distinct copy of a data structure or a memory image.
The in-NVM data structure also serves as the durable checkpoint. 
The challenge is to keep this checkpoint consistent as the structure is modified. 
After a crash, NVM state will consist of a mixture of the state at the beginning of the epoch, which must be kept, and modifications during the failed epoch, which must be discarded. 
The system must be able to distinguish between these two intermixed states and recover the one from the beginning of the epoch. 
One solution to this problem is to log the old memory value before each write. 
The log can be applied, in reversed application order, to roll back writes and to revert to the state at the beginning of the epoch. 
But logging itself requires care to ensure that the log reaches NVM before the structure is modified. 
Therefore, it again introduces the cost of cache-line flushes on the critical path. 

To solve the problem of fine-grained modifications to NVM, we introduce the new concept of an \emph{In-Cache-Line Log (InCLL)}.
An InCLL serves the same role as an undo log. 
But instead of using an external log, the InCLL is placed \emph{in the same cache line} as the data structure field being logged. 
InCLL relies on the Persistent Cache Store Order (PCSO) memory-ordering model of NVM (two writes to the same cache line reach NVM in program order) to ensure the ordering requirements of the log, without introducing cache flushes and delays. 

The main limitation of an InCLL is its limited capacity.
Since it resides in the same cache line as the data, an InCLL should be small and cannot handle all modifications.
If an object is modified multiple times during an epoch, InCLLs may be insufficient to provide crash recoverability.
In this case, our approach falls back on object-level logging.
After the entire object is logged, subsequent modifications in the epoch do not require additional actions. 
Together, the combination of InCLL and object-level logging drastically reduces the number of synchronous writes to NVM.\@

To validate this approach, we applied Fine-Grained Checkpointing into Masstree~\cite{Mao2012}, a cache-efficient data structure that combines a B+ tree and Trie.
We also implemented a durable memory allocator based on this approach. 
Measurements show that the overhead of our scheme is low and restart time is dramatically reduced.

The main contributions of this paper are:
\begin{itemize}
	\item Fine-Grained Checkpointing, a technique to ensure a consistent, quickly recoverable data structure in NVM after a system failure.
	\item In-Cache-Line Logging, a undo-logging technique that enables recovery of the state from the beginning of an epoch without requiring cache-line flushes in the normal case.
	\item Implementation of these techniques for the Masstree data structure, which made it persistent and demonstrated their application in a highly optimized system and their low (5.9-15.4\%) runtime overhead cost.
\end{itemize}

\section{Background}\label{sec-bqckground}

\subsection{Persistent Memory Ordering Model}

In this paper, we use the Persistent Cache Store Order (PCSO) memory ordering model for NVM~\cite{Cohen2017}. 
Cache lines are written back to NVM according to a computer's (unspecified) cache replacement policy, so we cannot assume any specific write-back behavior.
An application may explicitly force specific cache lines to be written to NVM, by using cache-line write-back instruction, such as the x64's {\tt clflushopt} or {\tt clwb}. 
These instructions are asynchronous, they only initiate a memory transfer but do not wait until data actually reaches NVM.\@
To ensure that a write-back completes, the application must issue a fence instruction, such as {\tt sfence}, which delays CPU execution until the outstanding write-back instructions finish.
Since this instruction waits until the data reaches NVM, it is far more expensive than a normal (cached) memory reference.

While ensuring the order in which writes to different cache lines reach NVM is expensive, ordering writes to the same cache line is essentially free. 
If two writes target the same cache line, the order in which they reach \emph{the cache} corresponds to the order in which they reach NVM.\@
Preserving the order of cache writes can be done with release memory ordering in C++11, which introduces a \emph{happens-before} relation between the writes~\cite{Manson2005, Boehm2008}.
On the x64 architecture, the release memory fence incurs \emph{no} runtime overhead and only limits the ability of a compiler to reorder writes.

Formally, given two writes \(X\) and \(W\), we say that \(X<_{p}W\) if \(X\) is written to persistent memory no later than \(W\).
\(X<_{hb}W\) is the standard \emph{happens before} relationship.
\(c(X)\) represents the cache line address \(X\) writes to. 
The following holds~\cite{Cohen2017}:
\begin{itemize}
	\item \(W<_{hb} \mbox{\emph{writeback}}(c(W))<_{hb}\mbox{\emph{fence}} <_{hb} X \Rightarrow \\ W<_{p}X\) (explicit flush).
	\item \(W<_{hb}X \wedge c(W) = c(X) \Rightarrow W<_{p}X\) (granularity).
\end{itemize}

Our InCLL technique relies on the second ordering guarantee, that if two writes target the same cache line, a happens-before relation is sufficient to ensure persistence ordering.

\subsection{Masstree Data Structure}\label{subsec-masstree-brief}

\def\perm{{\tt permutation}\xspace}
\def\vals{{\tt vals}\xspace}
\def\keys{{\tt keys}\xspace}

Masstree~\cite{Mao2012} is a production-quality ordered-set data structure that has been used to build in-memory databases such as Silo~\cite{Tu2013}. 
Masstree is a combination of a Trie and a B+ tree, implemented to carefully exploit caching, prefetching, optimistic navigation, and fine-grained locking. 
Below we sketch some details of Masstree that are necessary to understand our changes that make Masstree durable. 

Masstree uses two types of nodes: internal nodes and leaf nodes.
There are roughly an order of magnitude more leaf nodes then internal nodes and leaf nodes are modified much more frequently, so our checkpointing focuses on the leaf nodes. 
The number of items in an leaf node is a parameter of Masstree's implementation; the default implementation uses 15 keys and 15 pointers to values. 
The keys reside in the \keys{} array and the value pointers resides in the \vals{} array. 
Listing~\ref{listing-masstree-node} illustrates some details of a leaf node.
\begin{lstlisting}[caption={Masstree's node structure}, label={listing-masstree-node}, float]
class basenode; // lock, version, meta information
template <int width=15>
class leafnode : public basenode{
	basenode *parent, *prev, *next; 
	uint64_t permutation; // which key/vals are active
	keytype keys[width]; 
	valuetype *vals[width]; 	
	void remove(keytype key){
		int idx = find_idx(key); 
		remove_idx(&permutation, idx); 
	}
	void insert(keytype key, valuetype *val){
		int idx = insert_idx(&permutation); 
		keys[idx] = key; 
		vals[idx] = val;
	}
	void update(int idx, valuetype *val){
		vals[idx] = val; 
	}
};
\end{lstlisting}

The \perm{} field records valid key-value pairs, in other words which entries in the arrays are occupied. 
We can consider the \perm{} field to be a bitmap specifying whether an index is in use, although, in practice, it also orders entries as well.
Deleting a key-value pair from a node modifies only the \perm{} field. 
Inserting a new key-value pair modifies both the \perm{} field and an unused entry pair in the {\tt keys/vals} arrays. 
Updating an existing key modifies only the {\tt vals} entry. 

Masstree support splitting and merging of nodes, but these happen far less frequently than modification of leaf nodes. 
The full details of the Masstree algorithm are quite involved~\cite{Mao2012}, but are not necessary to understand this paper. 

\section{Overview}\label{sec-overview}

The main contributions of this paper are Fine-Grained Checkpointing and In-Cache-Line Logging, which we illustrate by show how to make Masstree durable.\footnote{Our code is online: https://github.com/epfl-vlsc/Incll}
The approach uses a combination of three techniques: fine-grained checkpoints, in-cache-line log, and external logging. 
This section sketches these three mechanisms, and \S~\ref{sec-algorithm} provides more detail.

\paragraph{Fine-Grained Checkpointing}
Execution is broken into \emph{epochs} --- our implementation uses 64\millisecond, the Masstree memory reclamation epoch, though longer or shorter intervals are feasible.
During an epoch, NVM contains a mixture of the memory state from the previous epoch and some --- but not all --- memory writes executed during the current epoch.
At the start of an epoch, the entire cache is flushed to NVM, ensuring that all modifications from the previous epoch are safely stored in NVM.\@
The cost of flushing the cache is low as its size is bounded and some modifications may have been written back to NVM during the previous epoch (\S~\ref{subsub-measure-global-flush}).

\paragraph{External Logging}
The external log is a standard undo log.
Under certain circumstances, when a Masstree node is modified during an epoch, the entire node is stored in the log so that subsequent modifications can be reversed.
To ensure persistence ordering, the log is written to NVM and an {\tt sfence} operation is issued before the node is modified. 

An external log is the standard tool for ensuring the consistency of durable data structures in NVM~\cite{Chakrabarti2014,Ching2017,Cohen2017,NVML,Hu2017, Kolli2016,Liu2017a,Volos2011}.
It is possible to log at different granularities: a word, an object, or a page. 
We choose object-level granularity, so that when a single word in a Masstree node is modified, the entire node is recorded in the external log. 
External logging's primary benefit is simplicity.
It ensures durability without requiring pervasive changes to the Masstree algorithm.
However, it may also have performance benefits since a node is only logged once, even if it is modified many times, as during merges or splits.

In our approach, we always use the external log for infrequent, these complex modifications.
In addition, changes to internal (non-leaf) nodes are infrequent, so they are also handled by the external log (\S~\ref{subsec-incll-internal}). 
Leaf nodes are logged only when required by the InCLL algorithm described below.

The external log is discarded after the cache is flushed at the start of an epoch since all of the logged changes will have been stored in NVM.\@

\paragraph{InCLL}
In-Cache-Line Logging (InCLL) is a technique for logging modifications to a node without waiting on NVM.\@
InCLL embeds an undo log inside the same cache lines as a Masstree leaf node. 
When a node is modified for the first time in an epoch, InCLL stores the old value of the modified field. 
Since the log resides in the same cache line as the data, no write backs or fences are required to ensure that the log reaches NVM with the modification. 

The primary benefit of InCLL is the low cost of logging. 
But, the capacity of each node's log is limited since an InCLL resides in the same cache line as the data.
Each log entry also reduces the number of Masstree array entries, which degrades the cache efficiency of the Masstree structure.
Our durable Masstree algorithm logs --- in a typical case --- only one or two modifications per node per epoch. 
If a leaf node is modified repeated, external logging is likely to be used. 

We find that the combination of a limited InCLL and an external log works extremely well. 
If Masstree updates during an epoch are random, most will access different nodes, and the external log will be infrequently used.
If, on the other hand, modifications are ordered, there may be many writes to the same node, in which case the external log, which only records the node once, will perform well.

\section{Durable Masstree}\label{sec-algorithm}

\def\curepoch{{\tt curEpoch}\xspace}
\def\permin{{\tt permutationInCLL}\xspace}
\def\epochnode{{\tt nodeEpoch}\xspace}
\def\lowerEpoch{{\tt lowNodeEpoch}\xspace}
\def\insAllowed{{\tt insAllowed}\xspace}
\def\logged{{\tt logged}\xspace}
\def\incllp{InCLL\(_p\)}
\def\incllv{InCLL\(_{1,2}\)}	
\def\true{{\tt true}\xspace}
\def\false{{\tt false}\xspace}

In our scheme, execution is partitioned into 64\millisecond\ epochs.
We use the {\tt wbinvd} instruction to flush the entire cache at the start of an epoch. 
Since Masstree uses epoch-based reclamation for allocating and de-allocating nodes, we reuse its mechanism and interval for our epochs as well.
Shorter intervals would raise the overhead cost of cache flushing (currently about 2\% (\S~\ref{subsub-measure-global-flush})) but reduce the number of updates that might be lost or need to be re-executed after a failure.

Epochs are assigned a monotonically increasing index, which is stored durably. 
With 64\millisecond epochs, a 32-bit index wraps after more than eight years.\footnote{If the data lasts longer, a background thread could run once every 8 years and reset all indices to zero.
Since the duration of graduate studies is less than 8 years, this feature is not in our current implementation.}
We also keep track of failed epochs.
During recovery, an epoch in which a crash occurred is added to the set of \emph{failed epoch}. 
Modification made during a failed epoch will not be visible after recovery. 

\subsection{InCLL Algorithm}\label{subsec-incll-algo}

In our durable Masstree algorithm, each node consists of 14 keys, 14 pointers to values, and two InCLLs.\footnote{One fewer key and pointer than the standard implementation.} 
Each InCLL requires 8 bytes, like a pointer. 
The InCLL entries are carefully cache aligned.
The first resides immediately before the 14-pointer array and the second resides immediately after the 14-pointer array. 
Hence, the first InCLL resides in the same cache line as pointers 0--6 and the second InCLL resides in the same cache line as pointers 7--14 (Figure~\ref{figure-node-structure}).
Each of these InCLL can record a single value modification per epoch. 

We use an additional InCLL for the \perm{} field. 
The \perm{} field records whether a location in the pointer array is occupied. 
This field is modified when a new key-value pair is inserted or removed.

The InCLL used to log the \perm{} field is denoted InCLL\(_p\), and the two InCLLs used to log values are denoted InCLL\(_1\) and InCLL\(_2\) for the first and second cache lines, respectively. 
The operation of \incllp{} differs from \incllv{}, as described below.

\subsubsection{InCLL Structure}\label{sec-incll-structure}
To understand the operation of the InCLL\(_p\) log, we describe how insert and delete operate. 
When a new key-value pair is inserted into a leaf node, an unoccupied location is found using the \perm{} field. 
The appropriate entry in the \keys{} array is set to the key and the corresponding entry in the \vals{} array is set to the value. 
Furthermore, the \perm{} field is updated to record that the entry is now occupied.
If there is no free entry, the node must be split, a case that is handled by external logging.

After a crash, a leaf node must be returned to the same state as the start of the current epoch.
There are a number of cases to consider.
If an entry was unoccupied at the beginning of the current epoch, there is no need to restore its \keys{} or \vals{} fields. 
Hence, the only field that must be logged is the Masstree \perm{} field (in \permin{}).
Even if multiple key-value pairs are inserted during an epoch, only the \perm{} field needs to be logged. 
Since this field is logged in InCLL\(_p\), there is no need to use the external log for multiple consecutive writes. 

Deletion in Masstree is similar to insertion.
Only the \perm{} field is modified to indicate that the entry for the key is now unoccupied, so no external logging is necessary in this case either.
Moreover, if a node is modified by inserting new key-value pairs and subsequently removing key-value pairs, then \incllp{} logging still suffices since restoring \perm{} leaves the node in its original state.

But it is not possible to simply log the \perm{} field in a mixed sequence of insertions and deletions.
An entry that is deleted might be overwritten by a subsequent insertion, which destroys the original key-value pair that should be restored after a crash. 
Thus, if a key-value pair is removed and, in the same epoch, a key-value pair is inserted into the same entry, then the entire node must be externally logged.
\incllp{} contains a boolean indicator (\insAllowed{}) that insertions do not require external logging. 
The indicator is initially \true{} and is set to \false{} during a delete. 

Since a data structure might be large, it is impractical to clear all InCLL entries when an epoch advances. 
The \incllp{} also records the epoch number in which the InCLL was used. 
During recovery, the log is applied (i.e., the \perm{} field is restored to its old value) only if the epoch number corresponds to a failed epoch. 

The last field of \incllp{} is a boolean (\logged{}) indicating that a node was logged to the external log. 
If the node was logged in the current epoch, no further logging is required.

\begin{figure*}
	\centering
	\includegraphics[width=0.85\textwidth]{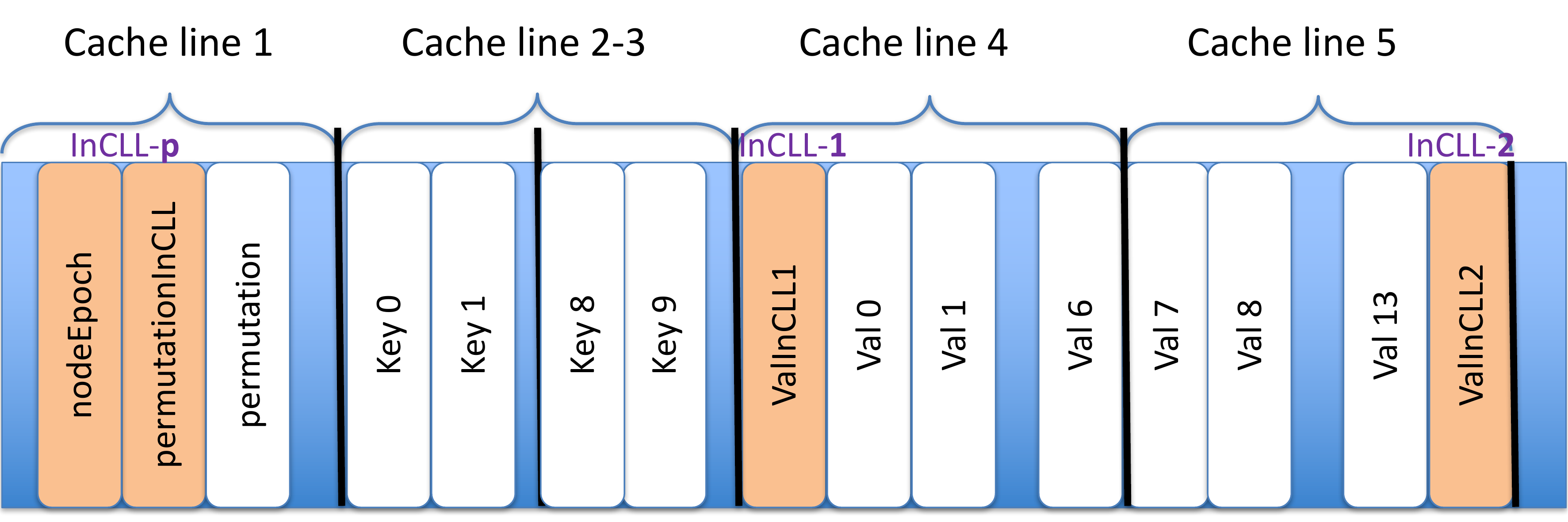}
	\caption{Durable Masstree leaf node. 
		InCLL are orange.
		\epochnode{} and \permin{} resides in the same cache line as \perm{}.
		Values have two InCLL entries, \incllv{}, located in their cache lines.}\label{figure-node-structure}
\end{figure*}

Overall, \incllp{} consists of four fields:
\begin{itemize}
	\item \epochnode{} stores the epoch number for the \incllp{}. 
	\item \permin{} stores the value of the \perm{} field at the beginning of the \epochnode{} epoch. 
	If the system crashes during this epoch, \perm{} is recovered from \permin{}. 
	\item \insAllowed{} controls if insertions are permitted to use the InCLL.\@ 
	\item \logged{} records if the node was logged in the external log during epoch \epochnode{}.
\end{itemize}
Code depicting the structure of a leaf node appears in Listing~\ref{listing-durable-masstree-node}.
It is also illustrated in Figure~\ref{figure-node-structure}. 

\begin{lstlisting}[caption={Durable Masstree's node structure}, label={listing-durable-masstree-node}, float]
class basenode; // lock, version, meta information
struct ValInCLL{
	long idx:4; 
	static const INVALIDIDX=-1; 
	long ptr:44; // 48 bits minus 4 least significant bits
	long lowNodeEpoch:16; // last 16 bits of the epoch; 
	ValInCLL(ptr, idx); 
	ValInCLL():ptr(nullptr),idx(INVALIDIDX); 
}; 
template <const int width=14>
class leafnode : public basenode{
	basenode *parent, *prev, *next; 
	uint62_t nodeEpoch; // @\incllp@ field
	bool logged, InsAllowed; // @\incllp@ fields
	uint64_t permutationInCLL; // @\incllp@ field
	uint64_t permutation; // which key/val are active
	keytype keys[width]; 
	alignas(64) struct {} ALIGN; // align to cache line
	ValInCLL InCLL1; // same cache line as vals[0..6] 
	valuetype *vals[width]; 	
	ValInCLL InCLL2; // same cache line as vals[7..13]
}; 
\end{lstlisting}

\subsubsection{InCLL Inserts and Deletes}

Next we consider inserting or deleting a key-value pair to a leaf node and show how \incllp{} is used (Listing~\ref{listing-masstree-ops}).
When a key-value pair is inserted or removed from a leaf node, the thread first checks whether \epochnode{} is equal to the current epoch (\curepoch{}).
If not equal, the current modification is the first time the node is modified in the current epoch. 
Then, the old (pre-modified) \perm{} value is saved in \permin{}, effectively logging its old value.
Afterward, \epochnode{} is set to the current epoch, \insAllowed{} is set to {\tt true}, and \logged{} to {\tt false}.

The persistence ordering of writing \permin{} and \epochnode{} is important. 
If setting \epochnode{} reaches NVM before \permin{}, recovery might fail. 
The problem is that if a failure occurs after the new epoch reaches NVM but before \permin{} reaches NVM, the recovery procedure (discussed below) will assume that the node was modified in the most recent epoch and must be recovered. 
Thus, it would incorrectly recover the node using the very old value of \permin{} (belonging to a previous epoch).

We use this persistence ordering:
Set \permin{} to the old (pre-modified) \perm{} value. 
Second, set \epochnode{} to the current epoch. 
Third, modify \perm{} to reflect the insertion or deletion of a key-value pair. 
The fields \insAllowed{} and \logged{} are semantically transient and do not require persistence ordering. 
This ordering ensures that the node can always be recovered. 
If only the first modification reaches NVM, the node is not recovered as \epochnode{} does not record a failed epoch. 
If the first and second modifications reach NVM, the node is recovered; in this case, both \perm{} and \permin{} represents the value at the beginning of the failed epoch. 
Recovery is not required, but also not harmful. 
If all three modifications reach NVM, the node is recovered correctly using \permin{}. 
These three fields reside in the same cache line, so ordering is ensured by a release memory fence, but not writing back or waiting for a full round-trip to NVM.\@

If \epochnode{} is equal to \curepoch{}, the node was already modified during the current epoch. 
As mentioned above, consecutive insertions or removals can use the InCLL.\@
The only case in which further action is need is when \logged{} is \false{} (that is, the node was not logged in the external log), the current operation is an insertion of a new key-value pair, and \insAllowed{} is \false{}. 
In this case, the node is logged in the external log before it is modified. 
Code for insertions and deletion appear in Listing~\ref{listing-masstree-ops}.

\begin{lstlisting}[caption={Durable Masstree operations}, label={listing-masstree-ops}, float]
void leafnode::InCLL(bool InCLLallowed, permInCLL,
                     valInCLL[2]){
	if(globalEpoch != nodeEpoch){
		isInsertionsAllowed = true; isLogged = false; 
		if(higher(globalEpoch) != higher(nodeEpoch)) 
			isLogged = logNode(); 
		if(!isLogged){
			permutationInCLL = permInCLL;
			InCLL1 = valInCLL[1]; 
			InCLL2 = valInCLL[2];
			// order writes
			atomic_thread_fence(memory_order_release);
		}
		nodeEpoch = globalEpoch; 
		InCLL[1,2].lowNodeEpoch = lower(nodeEpoch); 
	}
	else if(!isLogged && !InCLLallowed)
		isLogged = logNode(); 
	atomic_thread_fence(memory_order_release);
}
void leafnode::remove(keytype key){
	int idx = find_idx(key); 
	InCLL(true, permutation, ValInCLL(), ValInCLL()); 
	InsAllowed=false; 
	remove_idx(&permutation,idx); 
}
void leafnode::insert(typetype key, valuetype *val){
	int idx = insert_idx(&permutation); 
	InCLL(InsAllowed, permutation, ValInCLL(), ValInCLL()); 
	keys[idx] = key; 
	vals[idx] = val;
}
void leafnode::update(int idx, valuetype *val){
	ValInCLL InCLL = (idx<=6) ? InCLL1 : InCLL2; 
	InCLLallowed = (InCLL.idx == idx
	                || InCLL.idx == INVALIDIDX); 
	ValInCLL vc1 = ValInCLL (vals[idx], idx); 
	ValInCLL vc2 = ValInCLL(nullptr, INVALIDIDX);
	if(idx>=7) swap(vc1, vc2); 
	InCLL(InCLLallowed, permutation, vc1, vc2); 
	vals[idx] = val; 
}
\end{lstlisting}

\subsubsection{InCLL Update}

Updating the value of a key already in a leaf is slightly more complex than insertion or deletion. 
The two InCLLs embedded in the value array of a node, denoted \incllv{}, are used to log the old value when it is updated.
These InCLLs, unlike \incllp{}, require careful encoding to reduce their size.
\incllv{} can be used to log one of seven possible fields, so it contains an additional field that records the entry that was modified. 
Using two 64-bit words for \incllv{} would reduce the number of values in the array to 12 in two cache lines, incurring a performance penalty. 
Therefore, it is desirable to use a single word for \incllv{}.

To compact \incllv{}, we observe that the values stored by Masstree are pointers to the actual values. 
In current x64 architecture, pointers are represented in a canonical form in which only the lower 48 bits are used.\footnote{With future 5-level paging, we can fallback to external logging if the stored address is higher than $2^{48}$ or exploit stricter alignment restrictions.}
In a valid memory address, the upper 16 bits must equal the value of the 47th bit. 
In addition, all memory allocations are aligned on 16-byte boundaries, so the least significant four bits must be zero. 
We pack \incllv{} as follows. 
Bits 0--3 represent the index of the pointer that is logged; this field can represent seven values, so with 4 bits we can indicate all array entries (0--6 for InCLL\(_1\) and 7--13 for InCLL\(_2\)). 
Bits 4--47 hold the logged pointer. 
Bits 48--63 represent the lower 16 bits of the epoch in which the node was modified, denoted \lowerEpoch{}. 
We assume that the lower 16 bits of the current epoch can be combined with the higher 16 bits of \incllp{}'s \epochnode{} to produce the full epoch number in which the InCLL was used. 
During updates, we check if the difference between the current epoch and the \incllp{}'s \epochnode{}.
If 16 bits are insufficient to correctly encode the epoch, we fall back on the external log. 
This happens approximately once an hour (\(2^{16}\) epochs of 64\millisecond each). 

When the value of an existing key is updated, the thread first checks if \incllp's \epochnode{} is equal to \curepoch{}. 
If it is not, this is the first time the node is modified in the current epoch. 
The thread computes which InCLL must be used, depending on whether the modified entry's index is 6 or lower. 
The old value, the index, and the lower 16 bits of the epoch are encoded into a single word and stored in the appropriate InCLL.\@

If \incllp's \epochnode{} is equal to \curepoch{}, the node has been modified during the current epoch. 
But if the InCLL of the other cache line was used, it is still possible to use the unused InCLL.\@
In addition, if the pointer being modified was previously logged in the InCLL, there is no need to record it again. 
The latter is valuable when the keys are drawn from a skewed distribution. 
So, if some keys are popular and modified multiple times during an epoch, there is no need to use the external log. 
The external log is likely to be necessary if \emph{two} (or more) popular keys reside in the same cache line of a leaf node. 
Code for updates appear in Listing~\ref{listing-masstree-ops}. 

\subsection{External Logging}

We use the external log for modifications that are more complex or less common than these leaf updates. 
This has the benefit of requiring minimal changes to the Masstree code, while maintaining a relatively low overhead cost.
Splitting and merging of leaf nodes are infrequent and are handled by logging the affected nodes.
Also, all modifications to internal tree nodes are logged. 

The external log is also used whenever a modification cannot be handled by the InCLL.\@
If two values in the same cache line are modified in the same epoch, the external log is used. 
Similarly, if a key-value pair is removed and inserted at the same epoch, we also fall back on the external log. 

To reduce the cost of checking if an internal node was logged, we introduce an epoch number in each internal node that indicates that the node was log in a specific epoch.
A simple comparison against the current epoch number prevents multiple logging.
Our algorithm locks a node before logging to avoid races.
This provides an additional benefit during recovery; since a node appears at most once in the external log, there are no dependencies among log entries and they can be restored in parallel.
In contrast, standard undo logging has to be applied in a reversed application order, limiting concurrency in the recovery procedure.  

\subsection{Recovery}

After an abrupt crash, the durable Masstree is recovered as follows. 
First, the external log is applied before execution resumes. 
This is done by copying the contents of each node from the log to its corresponding node. 
As mentioned above, there are no dependencies among log entries, so it can be applied concurrently with minimal or no synchronization. 
Pseudo-code illustrating recovery appear in Listing~\ref{listing-masstree-recovery}. 

InCLLs must also be applied to recover nodes.
But unlike the external log, the InCLLs are embedded inside the durable Masstree nodes.
Applying all of them before the execution resumes would require a traversal of the entire tree, which would cause a long delay. 
To avoid this, the InCLL restores are applied lazily, during tree traversals.

When a thread attempts to access a node, it first checks if the node's \epochnode{} is less than the epoch number of the current execution. 
If it is, recovery is applied to the node before continuing with the access. 
To avoid concurrency races when multiple threads attempt to recover a node simultaneously, we use locking. 
However, it is not possible to use the leaf's lock. 
The problem is that the state of the lock is not preserved, so after a failure, it might be in a failed state. 
Therefore, attempting to lock the node could result in a deadlock, even if only a single thread is attempting to lock the node.
Instead, the system defines an array of (transient) locks for recovery. 
When a thread attempts to recover a node, it hashes the leaf address to find an appropriate recovery lock.
After acquiring the lock, the thread checks if the node's epoch is still lower than the first epoch in the current execution. 
If it is, the thread attempts to recover the node from the InCLL.\@
First, it checks if \epochnode{} is a failed epoch.
If so, the \perm{} field is recovered by copying the \permin{} field into it. 
Second, the thread reconstructs the epoch of \incllv{} by combining the lower 16 bits with the higher bits from \epochnode{}. 
If the resulting epoch number is a failed epoch, the index and value pointer are retrieved from the \incllv{} field and are applied to the appropriate location in the \vals{} array. 
Lastly, the node's InCLL is initialized to the first epoch in the current execution to indicate that the node does not need further recovery. 
Code illustrating InCLL lazy recovery appears in Listing~\ref{listing-masstree-recovery}. 

At the point when a leaf node is externally logged, it may have been modified and the changes logged in its InCLL.\@
Thus, the contents of the external log will not equal to the state of the node at the beginning of the failed epoch, and simply copying the log to the node is insufficient for correct recovery. 
After recovery using the external log, the InCLLs in the nodes must also be applied. 
To reduce recovery time, these restores are done lazily, on the first access to a node, similarly to non-externally logged nodes.

There is no need to flush cache lines during recovery.
If the system crashes before recovery is complete, it can be applied again.

\begin{lstlisting}[caption={Durable Masstree recovery}, label={listing-masstree-recovery}, float]
uint64_t currExecEpoch; // first epoch in current execution
lock recoveryLocks[K]; 
// before first access to durable Masstree
void durableMasstree::recovery(){
#  parallel for
	for each node L in external log do:
		memcpy(L->addr, L->content, L->size); 
}
// before first access to a leaf node
void leafnode::lazyNodeRecovery(){
	if(unlikely(nodeEpoch<currExecEpoch)){
		int idx = hash(this); 
		recoveryLocks[idx].acquire(); 
		if(nodeEpoch<currExecEpoch){
			nodeRecovery(); 
		}
		recoveryLocks[idx].release(); 
	}
}
void leafnode::nodeRecovery(){
	// InCLLp
	if(failedEpoch.find(nodeEpoch))
		permutation = permutationInCLL; 
	nodeEpoch = currExecEpoch; 
	// InCLL1
	uint64_t epoch = higher(nodeEpoch) | InCLL1.epoch; 
	if(failedEpoch.find(epoch))
		vals[InCLL1.idx] = InCLL.ptr; 
	// InCLL2
	epoch = higher(nodeEpoch) | InCLL2.epoch; 
	if(failedEpoch.find(epoch))
		vals[InCLL2.idx] = InCLL.ptr; 
	InCLL[1,2] = ValInCLL(nullptr, INVALIDIDX); 
	InCLL[1,2].epoch = lower(currExecEpoch); 
	basenode::initlock(); // might be in bad state after crash
}
\end{lstlisting}

\section{Durable Memory Allocation}\label{sec-mem-alloc}

\def\inclla{InCLL\(_n\)}

The Masstree data structure does not store actual values inside the tree.
Rather, Masstree holds a pointer to value buffers.
Clearly, these data buffer must be allocated in NVM so their contents remain after a crash. 
Allocating and deallocating these buffers are not strictly part of the Masstree algorithm.
Still, allocating a data buffer is typically required for every update operation on Masstree, so it is highly desirable to reduce the cost of such allocation by avoiding costly write backs and fences. 
As a second demonstration of Fine-Grained Checkpointing and In-Cache-Line Logging, we describe an allocation algorithm that uses them to avoid write backs and flushes during the critical path of allocation.

Our main observation is that an allocator is essentially a data structure (a set) that records free chunks of memory. 
Therefore, we can again use checkpointing for this structure, recovering the state of the allocator to the beginning of a failed epoch.
The InCLL technique can reduce the overhead of logging writes to the allocator's data structures. 

Our durable allocator uses a linked list of free objects, with a different list for each size class. 
When an object is allocated, it is popped from the appropriate list of free objects and inserted to Masstree. 
When an object is deallocated, it is added to the appropriate list of free objects and can be reused later (Masstree use epoch-based reclamation, so the memory becomes available only in the next epoch). 

To implement the list of free objects, it is sufficient to use a single next pointer per node. 
Thus, to implement a durable allocator, we protect this free pointer with an InCLL, denoted \inclla{}. 
The basic persistent allocator is simple, each object has a header of three words that fit into a cache line: the next pointer, the InCLL copy of the next pointer, and the epoch number. 
This design ensures that allocation never requires writing back to NVM or memory fences. 

Our allocator is based on Epoch-Based Reclamation (EBR), which allows a node to be allocated only if it was free at the start of an epoch (otherwise, concurrent threads might access the new data improperly). 
This property implies that there is no need to log the actual content of a buffer and no write backs are necessary. 
On recovery, the buffer is returned to its state at the beginning of a failed epoch, when the buffer is free and its contents are irrelevant. 

Our checkpointing technique is much simpler to use than other NVM systems' techniques for durable allocation. 
In these systems, a programmer has to specify explicitly that the buffer's contents must be written back to NVM before the structure operation is invoked. 
This is not the case for our system.
The programmer can simply write data into a buffer and insert it into the durable Masstree. 
At the start of the epoch, the contents of the buffer, together with the durable Masstree, will be written back to NVM when the entire cache is flushed. 

The drawback of our approach is that the allocator requires 24 bytes in the header of each object. 
Due to the 16 bytes alignment constraint, this results in a 32-byte header. 
But, we can reduce this overhead to only 16 bytes.

\subsection{Compacting InCLL for Memory Allocation}
\def\next{{\tt next}}
\def\nextin{{\tt nextInCLL}}
To ensure that the \emph{free list} (linked list of free nodes) return to the same state as the beginning of the epoch, each object needs a header with three fields: the current {\tt next} pointer; the value of the {\tt next} pointer at the beginning of the epoch, denoted {\tt nextInCLL}, used for logging; and a 32-bit epoch number. 
Since both {\tt next} and {\tt nextInCLL} are pointers in canonical form, the upper 16 bits of each can be computed from their 47th bit. 
Thus, the 32-bit epoch can be broken into two parts and encoded into \emph{both} next pointers, requiring only 16 bytes for all three fields. 

However, this idea is insufficient by itself. 
The problem is that a crash may happen after half of the epoch number reaches NVM but before the other half, leading to an incorrect recovery. 
Our solution is to encode a small counter at the two least significant bits of both \next{} and \nextin{}. 
This counter is incremented when the pointers are written in a new epoch and allows the recovery procedure to distinguish whether both pointers were fully written. 
If both pointers have the same counter value, the correct epoch number can be reconstructed by combining the 16 most significant bits of \next{} with the 16 most significant bits of \nextin{}. 
In this case, recovery proceeds similar to the \perm{} case (Listing~\ref{listing-masstree-recovery}). 
If, on the other hand, the pointers have different counter values, we know that a crash happen while the pointers were being modified, and the \next{} pointer must be recovered from the \nextin{} logged pointer.

\subsection{Correctness}

We tested the modified system by intentionally crashing it at random points, launching a new process, and checking that system's state matched the state at the beginning of the failed epoch.
We also used many unit tests to ensure that a cache line was left in its correct state.
We are currently developing a tool to help reason about the correctness of this type of system.

\section{Performance}\label{sec-measurements}

We implemented fine-grained checkpointing and InCLL as described above (INCLL) and measured it against the unmodified, transient Masstree (MT) and against an improved version of Masstree (MT+) that adopted two enhancements from INCLL, using a global barrier at each epoch and {\tt mmap}ing memory space for Masstree's pool allocator, rather than obtaining it through {\tt jemalloc}.
Without these two enhancements, INCLL performed slightly better than MT (Figure~\ref{figure-ycsb}).

All experiments were run on a server containing two Intel Xeon Gold 6132 (Skylake) processors, each with 14 cores and 28 hyperthreads, running at 2.6 GHz, with a 19.25 MB L3 cache.
The system contained 1.5 TB of DDR4-2666 RAM.\@
The operating system was Ubuntu Linux 16.04.5 LTS.\@
The code was compiled by g++ version 5.4.0. with the baseline makefile optimization level -O3.
Each experiment was executed 10 times, and we report the average time.
The standard deviation of the experiments ranged from 0.03\% to 0.08\%. 

Since NVM is not available, we allocate a file in {\tt /dev/shm} and mapped it to the application address space (DRAM). 
By default, we do not introduce artificial latencies for the cache flush or fence operations. 
However, we also measure the effect of higher NVM latencies by introducing artificial latency after the {\tt sfence} instructions since {\tt clflushopt} instructions are asynchronous.

Figure~\ref{figure-ycsb} reports the throughput of the three Masstree versions on different workloads.
Unless otherwise noted, the tree was initialized with 20 million entries and we ran with 8 threads.
Keys and values are 8-bytes long\footnote{Values are allocated in a 32-byte buffer containing additional Masstree fields.}.
The workload was generated by driver threads on the same machine to avoid network interference. 
For YCSB\_A (write heavy), the operation distribution was 50\% puts and 50\% reads, for YCSB\_B (read heavy) 5\% puts and 95\% reads, for YCSB\_C (read only) 100\% reads, and for YCSB\_E a read-only scan of 10 keys. 
We employ two key distributions. 
In \emph{uniform}, the keys are generated uniformly at random in the range between zero and 20M. 
In \emph{zipfian}, the keys are generated according to a zipfian distribution with a skew parameter of 0.99.
Keys are scrambled by computing a hash of their values, so that frequent keys do not (necessarily) appear in close proximity. 
We report the overall throughput (operations per second) of executing 1 million operations on each thread.

\def\figWidth{\columnwidth}
\begin{figure}
	\centering
	\includegraphics[width=\figWidth]{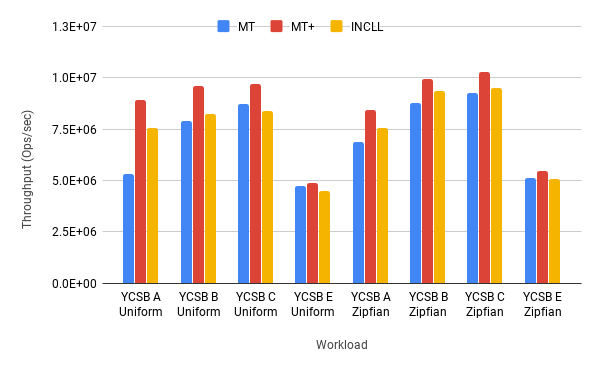}
	\caption{Throughput of baseline Masstree (MT), optimized Masstree (MT+), and our durable Masstree (INCC).}\label{figure-ycsb}
\end{figure}

\begin{figure}[]
	\centering
	\includegraphics[width=\figWidth]{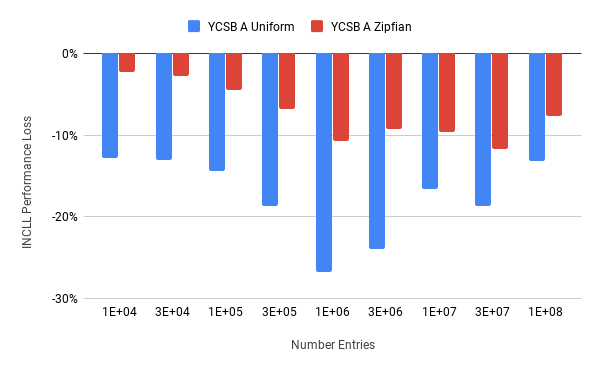}
	\caption{Overhead of INCLL over MT+ (INCL\_A benchmark) for varying tree size.}\label{figure-tree-size-perf}
\end{figure}

The optimized version (MT+) performed 2.4--68.5\% better than unmodified Masstree (MT). 
We use MT+ as the baseline for comparisons with the durable version (INCLL).
Durable Masstree performed 5.9--15.4\% slower than MT+, which reflects the cost of InCLL and periodic cache flushes. 
As expected, the write-intensive workload's (YCSB\_A) performance is reduced by a larger amount (10.3--15.4\%) than the read-light (5.9--13.9\%) or read-only workloads (7.9--13.5\%). 
The scan workload (YCSB\_E) is least affected by InCLL. 
The Zipfian workload performed better than the uniform workload in both systems, and it is less affected by InCLL (5.9--10.3\% vs. 7.8--15.4\%) because its skewed distribution means fewer nodes are accessed and consequently processor caching is more effective and more writes are logged (see discussion of Figure~\ref{figure-tree-size-perf}). 

INCLL increased the number of instructions executed 0.0--14.5\% (uniform) and 0.1--7.6\% (Zipfian). 
It also increased the number of L1 load and store references by a similar amount, but had less effect on the L1 cache miss rate. 
The number of LLC load references also increased by 7.9--16.3\%, but the number of L3 store references was less consistent (-28.7--27.5\%). 
For MT+, the LLC cache miss rate was high (25.8--39.9\% load, 25.1--99.6\% store), but the absolute number of misses is low (1.3--2.9M load, 31-666.8K store). 
INCLL had little effect on the L1 load miss rate (-5.1--14.2\%), but it reduced the LLC load miss rate by 42.0--95.2\%. 

\begin{figure}
	\centering
	\includegraphics[width=\figWidth]{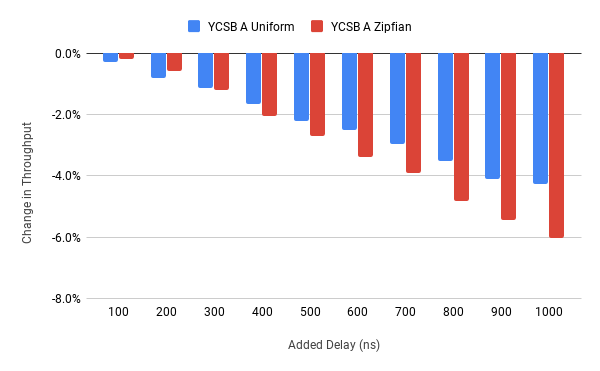}
	\caption{Effect of emulated latencies for cache write back on INCLL (baseline is INCLL with zero emulated latency).}\label{figure-added-latency}
\end{figure}

Figure~\ref{figure-added-latency} shows the effect of increasing the latency of flushing modified locations to NVM on the YCSB\_A write-heavy workload.
We introduced an additional delay of 100\nanosecond--1000\nanosecond\ after the {\tt sfence} operations.
The effect on NVM latency is small.
Even with an added latency of 1\microsecond, the performance of INCLL only decreased by 4.3\% for the uniform workload and 6.0\% for the Zipfian, compared to no emulated latencies. 
This small difference demonstrates that InCLL is able to avoid the full cost of flushing writing to NVM for most memory references. 

\begin{figure}
	\centering
	\includegraphics[width=\figWidth]{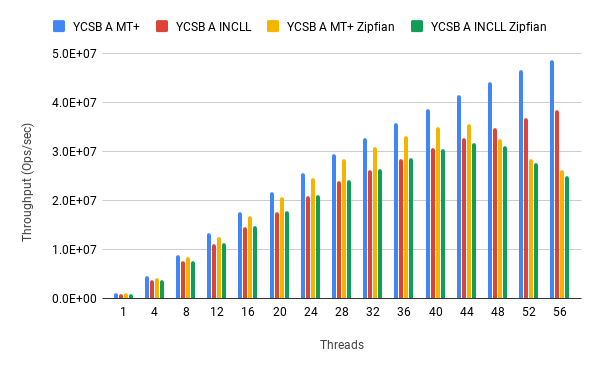}
	\caption{Throughput of INCLL and MT+ (INCL\_A benchmark) for different numbers of threads.}\label{figure-threads}
\end{figure}

Figure~\ref{figure-threads} depicts the performance of MT+ and INCLL over 1 to 56 threads.
The workload is again YCSB\_A.
The performance loss due to InCLL seemed unrelated to the number of threads, ranging from 14.6--21.3\% for uniform and 3.0--19.3\% for Zipfian. 
For the Zipfian workload, overall benchmark performance decreased at 44 threads in both system. 
In a larger tree (100M entries), however, the performance of all benchmarks increased monotonically with the number of threads and the performance loss due to InCLL was similar (10.7--15.3\% and 6.7--22.2\%, respectively). 

\begin{figure}[]
	\centering
	\includegraphics[width=\figWidth]{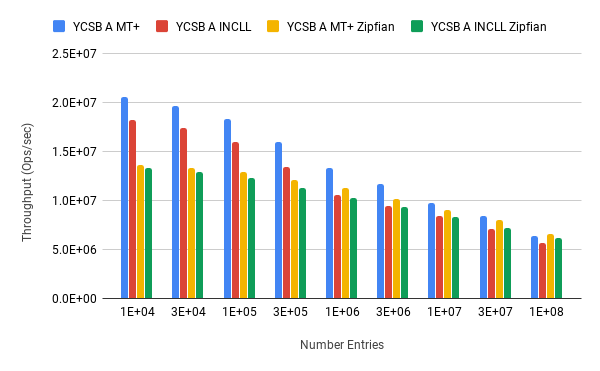}
	\caption{Throughput of INCLL and MT+ (INCL\_A benchmark) for varying tree size.}\label{figure-tree-size}
\end{figure}

Figure~\ref{figure-tree-size} depicts the performance of MT+ and INCLL for increasing the number of entries in the tree. 
The workload is again YCSB\_A. 
Both MT+ and INCLL were affected similarly by the increased tree size.
The performance on the uniform workload decreased 69\% for both MT+ and InCLL as the tree grew from 10K to 100M nodes, and the Zipfian workload performance decreased by approximate 50\%. 

\begin{figure*}
	\centering
	\includegraphics[width=\figWidth]{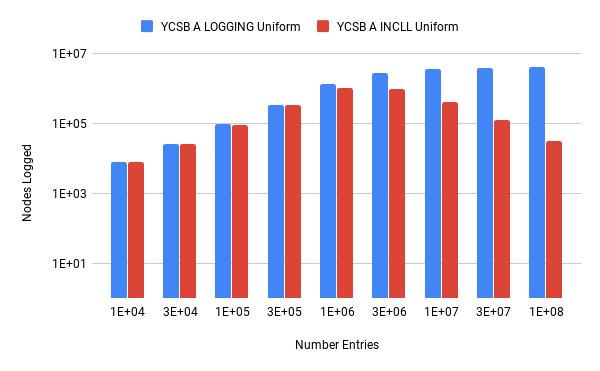}
	\includegraphics[width=\figWidth]{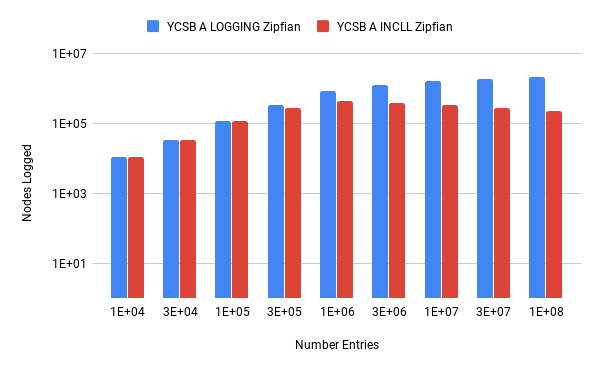}
	\caption{Number of logged nodes when InCLL logging is disabled (LOGGING) and with InCLL (INCLL).}\label{figure-nodes-logged}
\end{figure*}

The overhead for the uniform workload forms a parabola (Figure~\ref{figure-tree-size-perf}), with a lower overheads for small and large trees.
To understand this phenomena, we measured the effectiveness of the external log and InCLL.\@
Figure~\ref{figure-nodes-logged} show the number of nodes logged for the YCSB\_A workload when InCLL logging is disabled (LOGGING) and in the normal operating mode (INCLL).
As can be seen, for both uniform and zipfian distributions, the number of logged nodes increases sharply (logarithmic plots) until the tree reaches 1--3M nodes.
After that point, the uniform distribution levels off without InCLL and declines rapidly with InCLL, and the zipfian distribution grows slowly without InCLL and declines slowly with InCLL.\@

A tree of 10K keys contains approximately 1K nodes, many of which are modified frequently by the approximately 80K operations during a typical epoch.
A node modified a second time is usually recorded in the external log (\S~\ref{sec-incll-structure}).
However, no logging is needed for the rest of the epoch, so subsequent modifications incur no overhead.
For a tree with 100M keys, however, overhead is low for a different reason. 
When the tree is very large and keys are chosen uniformly at random, a node has a low probability of being modified and a lower probability of being modified twice.
Thus, most of its modifications are logged by InCLL, not external logging.
The Zipfian distribution, which has a higher locality of reference and more nodes access twice or more, does not benefit as much from the InCLL and its number of nodes logged continues to increase as the tree grows.
In the middle, when the tree contains 1M--3M entries, the probability that a node is modified twice or more is relatively large. 
This entails external logging, but since the number of operations on a given node is likely to be low, the overhead of this logging is unlikely be amortized over a series of operations, as in smaller trees. 
Despite the relatively high overhead for trees in this range, the heavy-write benchmark ran at most 27\% slower than MT+. 

\begin{figure*}
	\centering
	\includegraphics[width=\figWidth]{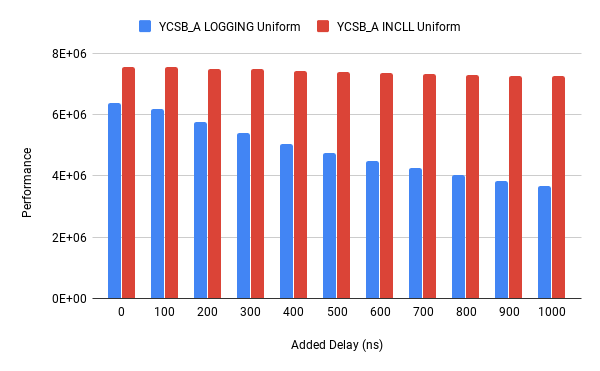}
	\includegraphics[width=\figWidth]{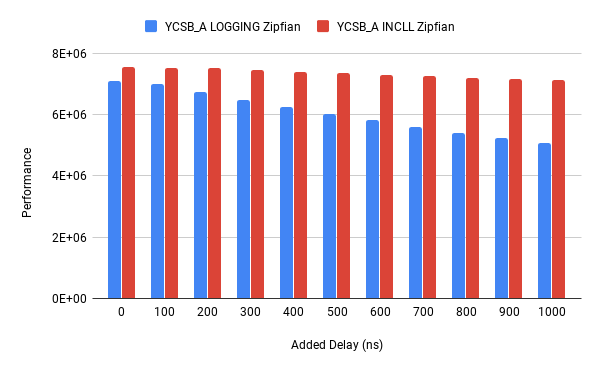}
	\caption{Performance with emulated latencies for cache write back when InCLL logging is disabled (LOGGING) and with InCLL (INCLL).}\label{figure-delay-logging}
\end{figure*}

Figure~\ref{figure-delay-logging} show how performance changes for both InCLL (INCLL) and only logging (LOGGING) as the latency of flushing modified locations to NVM increases on the YCSB\_A workload.
With InCLL, the performance decreases only 4.1\% (uniform) and 5.7\% (zipfian) as the latency of this operation increases to 1\microsecond.
With only logging, the performance decreases 42.5\% (uniform) and 28.5\% (zipfian) over this range.
InCLL greatly reduces the number of cache lines that must be flushed, which becomes increasingly beneficial as the latency of this operation increases.

\subsection{InCLL for Internal Nodes}\label{subsec-incll-internal}

An initial version of our system applied InCLL to internal nodes as well. 
This significantly reduced the number of internal nodes that were logged. 
However, it did not improve performance appreciable since many more leaf nodes are logged.
Furthermore, InCLL reduced the width of internal nodes, resulting in slower tree traversals. 
Overall, applying InCLL to all nodes resulted in lower performance.

\subsection{Global Flush}\label{subsub-measure-global-flush}
We measured the cost of a global cache flush using the three workloads.
The instruction that flushes the entire cache, {\tt wbinvd}, is a privileged instruction that can only be executed by the kernel. 
We measured user-space overhead: the time from the syscall until the operation returned to user space. 
In all measured cases, the cost was 1.38--1.39\millisecond with a variance of 6-12\%. 
Since the flushes are executed once every 64\millisecond, the total cost of this operation is 2.2\%. 

\subsection{Recovery Time}\label{subsec-measure-recovery}
To measure recovery, we intentionally crashed the system immediately before starting a new epoch.
This is a worst-case scenario for the number of nodes recorded in the external log. 
The workload is write-heavy (50\% writes) and the tree was 1M entries (worst-case scenario for InCLL). 
We found that 84K nodes were recorded during the epoch.  
Applying these log entries required approximately 15\millisecond. 
As expected, recovery is fast, even in a worst-case scenarios, primarily because of the short epoch duration.

\section{Related Work}\label{sec-related}

There have been many attempts to create efficient durable indexes in NVM.\@
wB+-Trees~\cite{CJ+15} are a tree designed to reduce the number of writes to NVM by using unoccupied leaf entries for insertions, in a manner somewhat similar to the permutation field in Masstree. 
However, wB+-Trees still require at least two write backs and fences per update. 
NV-Tree~\cite{Yang2015} uses an append-only strategy to reduce the overhead of writing to NVM.\@
However, every split requires reconstructing the path of internal nodes, increasing overhead. 
Update operations still require two write backs and fences.
FPTree~\cite{Oukid2016} reduces the overhead of NVM writes by storing only leaf nodes in NVM.\@
Internal nodes are stored in DRAM and rebuilt during recovery. 
However, rebuilding the tree increases restart time significantly. 
In addition, it requires three write backs and fences per update operation. 
WORT~\cite{Lee2017} is a radix tree designed for NV, which attempts to reduce the number of writes to NVM.\@
However, it still requires two write backs and fences per update operation. 
BzTree~\cite{Arulraj2018} uses a lock-free persistent multi-word CAS operation (PMwCAS) to implement a durable B+ tree. 
They do not report the number of write backs and fences (which might vary due to concurrency races), but at least two are needed for each PMwCAS.\@
Insertions require at least two PMwCAS, so at least four persistent fences are necessary. 
\def\dali{Dal{\'{i}}}
\dali~\cite{Nawab2017} is a durable, nonblocking hash map based on globally flushing the cache. 
Each bucket follows an append-only strategy, so updating an existing value allocates a new node and appends it to the corresponding hash bucket. 
Therefore, \dali{} makes less efficient use of the cache than Masstree.
Memory reclamation relies on a garbage collection-like algorithm during recovery, which makes recovery very expensive.

Most programming interfaces to NVM are either transactional memory or locks. 
Mnemosyne~\cite{Volos2011} was the first system that used a software transactional memory system for NVM.
It is based on a durable redo log whose implementation is from TinySTM~\cite{Felber2008}. 
PMDK~\cite{NVML} is a library, provided by Intel, that uses an undo log to provide durable transactions as an NVM interface.
Atlas~\cite{Chakrabarti2014} uses locks to delimit uninterruptible durable regions. 
It uses a durable undo log to roll back unfinished atomic regions after a failure.
The InCLL programming model is more complex, as it tailors logging to the semantics of a data structure, but it achieves far better performance than general-purpose approaches.

Numerous papers have tried to improve transaction performance without changing the programming model.
None come close to the low overhead of InCLL.\@
Kolli et al.~\cite{Kolli2016} pipelined multiple stages to reduce the number of write backs and fences. 
NVThread~\cite{Ching2017} improved the lock-based durable section with a redo-log. 
Different threads are spawned as different processes, giving each thread a fast, hardware mediated view of its local modification. 
LSNVMM~\cite{Hu2017} improved the performance of durable transactions by avoiding replication of data in a log and the original location. 
Kamino-TX~\cite{Memaripour2017} avoids slow lookups in the redo log by applying modifications to a DRAM copy. 
These modifications are propagated lazily to NVM.\@
DudeTM~\cite{Liu2017a} decouples transactional concurrency control from the durability mechanism. 
Concurrency control is performed on a DRAM copy and produces a redo log, which is applied to NVM in the background.
The latter two systems avoid write backs and fences on the critical path but suffer from a long restart delay due to the cost of copying the NVM structure to DRAM.\@

The main challenge that needs to be solved by a persistent memory allocator~\cite{DBLP:conf/vldb/SchwalbBFDP15, Bhandari2016, Nawab2017, Coburn2011}
is inconsistency between the allocator's metadata and the application's data structures.
If a buffer was allocated but the system crashed before it was linked to the persistent structure, it is a persistent memory leak.
Schwalb et al.~\cite{DBLP:conf/vldb/SchwalbBFDP15} broke allocation into two steps, reserve and activate, where each flushes data to persistent memory.
A crash after the reserve step is rolled back by the system. 
The NV-heap~\cite{Coburn2011} system implements allocation, automatic garbage collection, reference counting, and pointer assignments as simple, fixed-size ACID transactions using a persistent redo log.
Makalu~\cite{Bhandari2016} uses a conservative garbage collector to recover unreachable pointers.
Thus, no writes back are required during allocation. 
It has a slow restart time, due to its need to traverse a potentially unbounded amount of memory before an application restarts.

\section{Discussion}

Fine-Grain Checkpointing is not an exact replacement for transactions and InCLLs are not a general-purpose substitute for logs.
Using InCLLs require detailed understanding of data access patterns and cache-line boundaries.
Checkpointing requires less program annotation than transactions since a developer only needs to ensure data structure consistency at infrequent epoch boundaries, but this does not alleviate the complexity introduced by InCLL and does not guarantee immediate durability.
Their combination provides a powerful tool for experts, such as library developer, for reducing the cost of durability, albeit at the cost of additional programming complexity. 

Analogous to cache-efficient or concurrent data structures, we believe that efficient, recoverable structures cannot be created with one-size-fits-all techniques.
Achieving high performance require code that is data-structure and architecture specific.
In this paper, we use Masstree as an example to demonstrate how InCLLs make a highly optimized structure durable.
The amount of effort was approximately 2 person months without prior knowledge of Masstree.

Other pointer-based structures could benefit from these techniques.
Currently, achieving good results requires careful reasoning about the characteristic of the specific structures (e.g. \S~\ref{sec-algorithm}).
We are aware there are still many open questions about how to apply this technique in other contexts --- e.g., values that span a cache line, objects with less sharply defined update patterns, or more complex data structure manipulations --- and are investigating more general solutions.

\section{Conclusion}\label{sec-conclusion}

In this paper, we present Fine-Grained Checkpointing and In-Cache-Line Logging.
The former periodically flushes the cache to NVM, thereby persisting everything.
The latter embeds an undo log inside the cache lines in a data structure, thus enabling fast logging to undo writes from partially executed epochs.
We transformed the Masstree data structure to be durable using these techniques and an external object-granularity log to handle complex and infrequently written structure operations.
The combination of these techniques guarantees durability while introducing only a moderate performance overhead.

\bibliographystyle{plain}
\bibliography{bib/NVM,bib/OODBMS,bib/ManagedRecovery,bib/BoundPersistentFlush,bib/NVMIndex,bib/MemoryModel}

\end{document}